\documentclass{optica-article}

\journal{opticajournal}
\preto {\abstractkeywords} {\nolinenumbers}
\usepackage{graphicx}
\usepackage{multirow}
\usepackage{booktabs}
\usepackage{appendix}
\usepackage{array}
\usepackage{braket}
\usepackage{amsmath}
\usepackage[marginal]{footmisc}
\usepackage{bm,color,bbm}

\usepackage{float}
\newcolumntype{.}{D{.}{.}{-1}}
\usepackage{amsopn}

\articletype{Research Article}
\usepackage{hyperref}	
\hypersetup{
	colorlinks = true,
	citecolor = {blue},
	linkcolor = {red}
}
\usepackage{lineno}

\begin{document}

\title{Resource-efficient quantum key distribution with integrated silicon photonics}

\author{Kejin Wei,\authormark{1,4,$\ast$}  Xiao Hu,\authormark{2,4} \author{Yongqiang Du,\authormark{1} Xin Hua,\authormark{2,3} Zhengeng Zhao, \authormark{1} Ye Chen, \authormark{1} Chunfeng Huang, \authormark{1} Xi Xiao, \authormark{2,3,$\dagger$} }}

\address{\authormark{1} Guangxi Key Laboratory for Relativistic Astrophysics, School of Physical Science and Technology, Guangxi University, Nanning 530004, China\\
	\authormark{2} National Information Optoelectronics Innovation Center (NOEIC), 430074, Wuhan, China\\
	\authormark{3} State Key Laboratory of Optical Communication Technologies and Networks, China Information and Communication Technologies Group Corporation (CICT), 430074, Wuhan, China\\
	\authormark{4} These authors contributed equally to this paper.}



\email{\authormark{$\ast$}kjwei@gxu.edu.cn} 
\email{\authormark{$\dagger$}xxiao@wri.com.cn} 



\begin{abstract*}	
	Integrated photonics provides a promising platform for quantum key distribution (QKD) system in terms of miniaturization, robustness and scalability. Tremendous QKD works based on integrated photonics have been reported. Nonetheless, most current chip-based QKD implementations require additional off-chip hardware to demodulate quantum states or perform auxiliary tasks such as time synchronization and polarization basis tracking. Here, we report a  demonstration of resource-efficient chip-based BB84 QKD with a silicon-based encoder and decoder. In our scheme, the time synchronization and polarization compensation are implemented relying on the preparation and measurement of the quantum states generated by on-chip devices, thus no need additional hardware.  The experimental tests show that our scheme is highly stable with a low intrinsic QBER of $0.50\pm 0.02\%$ in a 6-h continuous run.  Furthermore, over a commercial fiber channel up to 150 km, the system enables realizing secure key distribution at a rate of 866 bps. Our demonstration paves the  way for low-cost, wafer-scale manufactured QKD system.	
\end{abstract*}


\section{Introduction}
Quantum key distribution (QKD) has been believed as a promising tool   providing theoretic information security without any restrict on the eavesdropper's computing power. Since the first QKD scheme, named BB84 protocol~\cite{1984bennett}, was proposed  by Bennett and Brassard in 1984, QKD has attracted a lot of interests and achieved a rapid development. 
So far, several QKD theoretical schemes have been proposed to improve performance or enhance security~\cite{2012Braunstein-MDI,2012Lo,2014Sasaki,2018Lucamarini,2018Maxiongfeng,2019Curty-TF,2018wangxiangbin,2021Schwonnek-DI}. Experimentally,  QKD has been widely implemented towards long distance~\cite{2018Boaron2}, GHz repetition rate~\cite{2018Boaron1,2020Gru-5GHz}, field test~\cite{2021Avesani-field}, large scale network~\cite{2019Dynes-network,2021Chenyuao-network}. The recording transmission distance of QKD was pushed up to 830-km installed fiber~\cite{2022wang-TF}. Recently, important breakthroughs were achieved in the demonstration of device-independent QKD~\cite{2022-Liu-DI,2022-Nadlinger-Di,2022-Zhang-DI}.   Recent researches have focused on  making QKD systems simpler, more compact and lower-cost  for widespread deployment~\cite{2020Agnesi,2021Madi}.   

Integrated photonics is a natural solution to realize miniature QKD~\cite{2016Silvertone,2019Wang}.
In particular, QKD chips  relying on well-established integrated
techniques have been fabricated to realize key QKD devices~\cite{raffaelli2018homodyne,2021Gyger-chipdetector,2021Beutel,2022Beutel-chip,2022Ye-chip}. The reliability of chip-based QKD devices has also been demonstrated by implementing several QKD protocols, including decoy-state BB84~\cite{2016Machip,2017Sibson-chip,bunandar2018metropolitan,paraiso2019modulator,avesani2019full,2021Zhang-chip,2022Li-PLC,2022Beutel-chip}, high dimension~\cite{ding2017high,2022Zahidy}, continuous variable~\cite{zhang2019integrated}, measurement-device-independent QKD~\cite{2020cao,2020Wei,2021Li-Chip,2021Zheng-MDI-CHIP}.  These achievements open the way to implant QKD into telecommunication networks. See Ref.~\cite{2022Liu-Review} for a recent review for advance in chip-based QKD.

\begin{figure*}[!hbt]
	\centering
	\includegraphics[width=1\linewidth]{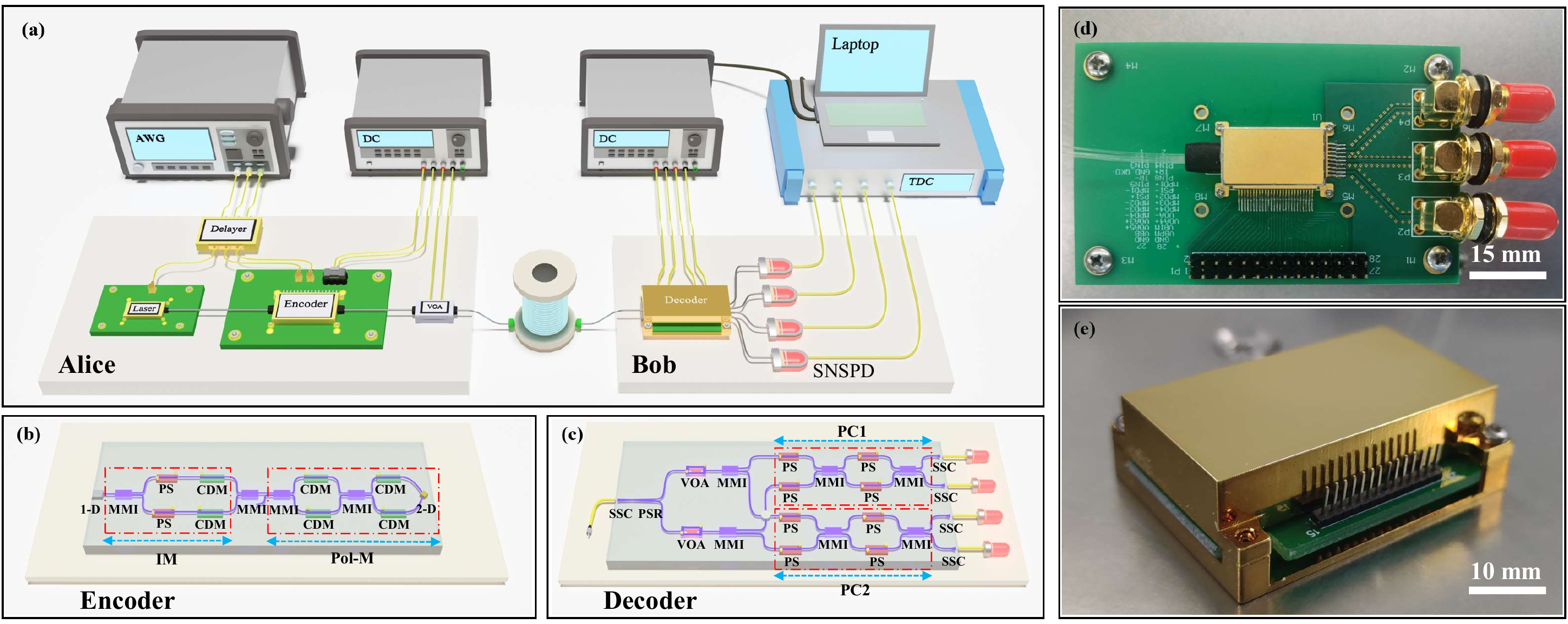}
	\caption{Silicon-based QKD system: (a) Schematic of the QKD setup. Laser, a laser diode; Encoder, a silicon chip integrates a intensity modulator and a polarization state modulator; VOA, variable optical attenuator; Decoder, polarization state demodulation chip;  SNSPD, superconducting nanowire single photon detectors; TDC, time digital convert. AWG, arbitrary waveform generator; DC, programmable DC power supply; Delayer, a home-made time delay generator. (b) Schematic of the encoder.  MMI, multimode interferometer; PS, thermo-phase shifter; CDM, carrier depletion modulator;  1-D, one dimensional grating coupler; 2-D, two dimensional grating coupler; IM, intensity modulator;  Pol-M, polarization modulator. (c) Schematic of the decoder. SSC, spot-size converter; PSR, polarization splitter-rotator. (d) Picture of the packaged encoder chip soldered to the external control board. (e) Picture of the packaged decoder chip soldered to the external control board. }
	
	\label{setup}
\end{figure*}

Until now, however, chip-based QKD experiments have  two substantial limitations. Firstly, although remarkable advances in quantum integrated devices enable chip-based demonstration of QKD, it  requires additional off-chip hardware to perform auxiliary tasks in the  process. For example, in recent reported multichip QKD system~\cite{2021Paraiso-chip}, the time synchronization between Alice and Bob is realized using an additional classical light synchronization setup. In the photonics polarization decoder presented in Ref.~\cite{2021Zhang-chip}, a bulk fiber-based polarization controller is used to track polarization basis as well as compensate polarization drift.

Secondly, despite that standard silicon photonic fabrication provides a feasible way to realize mutlifuctional devices for manipulation of quantum states, it is hard to fabricate some integrated components for quantum applications. For example, some prior works involving chip-based polarization-encoding system  still use discrete optics devices to demodulate the quantum states  due to the difficulty of achieving integrated polarization manipulation devices~\cite{2016Machip,bunandar2018metropolitan,avesani2019full,2020Wei}.  
Hence, building a fully deployed QKD system  is still challenging.

In this work, we present a resource-efficient chip-based polarization-encoding QKD scheme in which key distribution and all auxiliary tasks are realized using the same chip devices.  The scheme is implemented  using  chip-based encoder and decoder that are manufactured by standard Si photonic platform. The time synchronization and polarization drift compensation are performed exploiting a qubit-based method, which is relied on the quantum states generated by the chip devices.  The scheme experimentally shows an average QBER of $0.50\pm 0.02\%$ and high stability in the 6-h continuous operation.  Furthermore, we successfully generate secure bits over  different fiber channel lengths, up to 150-km fiber spool with a finite-key rate of 866 bps. This study  paves the way for low-cost, wafer-scale manufactured QKD system and provides a promising solution for making chip-based systems simple.


\section{Setup}

Our chip-based BB84 QKD system is schematically shown in Fig.~\ref{setup}. Alice prepares decoy-state BB84 polarization states on a chip-based transmitter and distributes the states to Bob, who possesses a chip-based receiver, via an optical fiber link.

Alice generates low-jitter phase-randomized light pulses at a repetition rate of 50 MHz and a center wavelength of 1550~nm. The generated pulses are coupled into a Si photonic encoder chip (see Fig.~\ref{setup}(b)) which integrates together an intensity modulator (IM) and a path-to-polarization modulator (Pol-M). The components consist of  interferometric structures which exploit standard building blocks offered by  the commercial fabrication foundry.  The multi-mode interference (MMI) couplers act as symmetric beam splitters, and the thermo-optics phase shifters (PSs) with $\sim$5 KHz bandwidth, and carrier-depletion modulators (CDMs) with $\sim$5 GHz bandwidth act as phase modulators.

The first interferometric structure, realizing a Mach-Zehnder interferometer (MZI), is used to realize an IM. The IM is used to generate  different intensities, one intensity $\mu$ for signal state and another one $\nu$ for decoy state. The
ratio $\mu/\nu$ is set by both  DC-bias of the PSs and  the RF signals sent to the CDMs of the MZI.  

The output of IM is connected to Pol-M which functions by means of path-to-polarization converter. The structure combines an inner MZI with two external CDMs, ending in a 2 dimensional grating coupler (2-D). The 2-D, acting as a polarization rotator combiner, converts the path-encoding information at each arm of the MZI into polarization-encoding information at the output. The Pol-M enables preparing the four BB84 polarization states, $\left| \psi  \right\rangle  =( \left| H \right\rangle  + {e^{i\theta }}\left| V \right\rangle)/\sqrt{2} ,\theta  \in \{ 0,{\pi  \mathord{\left/
		{\vphantom {\pi  2}} \right.
		\kern-\nulldelimiterspace} 2},\pi ,{{3\pi } \mathord{\left/
		{\vphantom {{3\pi } 2}} \right.
		\kern-\nulldelimiterspace} 2}\}$, where $\theta\in\{0, \pi\}$ ($\theta\in\{\pi/2, 3\pi/2\}$) represents the states in $Z$ ($X$) basis.  More details about chip design as well as work principle can be found in Ref. \cite{avesani2019full}

All the operation of Alice's devices, including triggering the laser, applying voltage on the IM and Pol-M,   are controlled by  arbitrary waveform generators (AWGs) and  home-made time delay generators (Delayer).    Alice's encoding pulses are sent to silicon-based measurement setup in Bob.  

The transmitted signals are coupled into Bob's decoder chip (see Fig.~\ref{setup}(c)) via spot-size converter (SSC). The decoder chip then converts the polarization-encoding photons to on-chip path encoding  via polarization splitter-rotator (PSR). Therefore, these photons are passively choose the measurement bases of $Z$ or $X$ via  a symmetric multi-mode interferences (MMI).  The quantum measurement in $Z (X)$ base on constituted by a polarization controller measurement state PC1 (PC2). Each PC contains two MMIs and four thermal phase shifters (PSs). By fine-tunning the applied voltages, supplied by a DC power supply,  on PSs, the PC1 and PC2 constitute the quantum measurement in $Z$- and $X$-basis, respectively. Further information about the principle of polarization state analysis by the decoder chip for the BB84 protocol can be found in Ref.~\cite{2022Du}

After configuring the measurement basis in PCs, the photons are detected by  four commercial superconducting nanowire single photon detectors (SNSPDs,  Photon Technology Co., Ltd. ). The SNSPDs is cooled down to 2.1 K and have an detection efficiency of around 75\%, dead time of $\sim$40 ns, time jitter of $\sim$70 ps and dark counts of $\sim$25 Hz. The detection events are registered using a high-speed time tagger (Timetagger20, Swabian Instruments). The local clock reference for Bob are generated by using internal clock in TDC. A laptop then reads the TDC data and processes  it for time synchronization and  polarization drift compensation. The encoder is triggered by a programmable DC power supply.

The encoder and decoder chips are fabricated using  stand silicon photonics foundry services with a in-house design. The encoder and decoder have a footprint size of $6\times3$~mm$^2$  and  $1.6\times$1.7~mm$^2$, respectively. The encoder chip is butterfly packaged with a total volume of 20 $\times$ 11 $\times$ 5 mm$^3$   and then  soldered to standard $9\times7$~cm$^2$ printed circuit board, as shown in Fig.~\ref{setup}(d). The decoder chip is packaged using chip-in-board assembly with a total volume of 3.95 $\times$ 2.19 $\times$
0.90 cm$^3$, as shown in Fig.~\ref{setup}(e).  With dedicated layout, the chip is easily assembled by  commercial foundry, providing a low-cost, portable, stable and miniaturized device.  Details for the encoder and decoder chip designs and fabrications can be found in~\ref{chip-design}.

\subsection{Qubit-based time synchronization and polarization compensation}\label{qubit-method}

Different from the conventional optical or electrical  time synchronization methods, we apply a qubit-based method to synchronize two remote users' clocks. This method enables synchronizing the clocks using the transmitted quantum signal states.  This means that our setup does not require a synchronization subsystem, which  normally consists of a pulsed laser and a photonic detector, as utilized in pioneering works~\cite{2019Bacco,2021Geng,2023Li-Chip}.  We remark that removing the synchronization subsystem is extremely crucial for building a fully chip-based QKD system since the development of a on-chip laser source remains a great challenge on Si-photonics research~\cite{2022Han-Laser}.

The main goal of the time synchronization is to recovery the specific frequency of the qubits   arriving at the detectors and the time-offset of corresponding qubits between Alice and Bob. The main idea of our algorithm is similar to a recently proposed $Qubit4Sync$ algorithm~\cite{2020Calderaro} except that the method of recovering the absolute time-offset. instead of implementing public periodic-correlation codes in the first $L$ states, we divide those codes into single bits, then periodically and cyclically embed them in prepared states. The detailed description of our synchronization method is presented in~\ref{time-syn}. Here, we describe the main features of our algorithm as follow.

Our approach first computes the frequency
from the time-of-arrival measurements based on fast Fourier transform. To recover the absolute time-offset, Alice prepares a random qubit string which is periodically and cyclically embed with several  public known bits form periodic-correlation codes with a length of $L$. Then each bit, according to a preset ratio $M$, is sequentially inserted into random encode bits.  For example, suppose that $M=9:1$, every 10 bits contain 9 random bits are embed with a single bit from  periodic-correlation codes of length $L$.  When the $L$ bits are consumed, the periodic-correlation codes are cyclically reused.
By correlating these periodic-correlation codes with the one received by Bob, it is possible
to distinguish which state received by Bob is the first one sent by
Alice. The absolute time-offest of the first qubit can be obviously derived. This is the typical frame-synchronization
technique used in digital communication systems~\cite{2000.Adriaan.J.IEEE}.

Our algorithm periodically inserts  the synchronization string in the random data stream so that Bob can execute the qubit synchronization algorithm at any time and quickly re-establish synchronization when the system is out of step.   This allows the system to continuously accumulate bit string for a long time up to $6$ hours. 

After recovery time synchronization, we can further perform polarization compensation using a shared qubit string. We note that,  due to the environmental disturbance, the polarization state of transmitted photon is rotating unpredictably along fiber channel. This polarization variation would leads to a mismatch between Alice and Bob's reference frames and resulting in a high QBER and low secure key rate. To eliminate it, a polarization compensation subsystem, which typically includes an auxiliary laser and several electronic polarization controllers,  must be utilized. 

In our scheme, since the POVMs in the $Z$ and $X$ bases are set by actively tuning PSs at the stages of PC1 or PC2. which enables the polarization basis tracking and hence allows our system to compensate polarization rotation. 

Here, similar to the work in~\cite{2017Ding-polarization,2020Agnesi}, we use qubit-based polarization compensation method that employs a shared qubit string. The shared qubit string are generated using the chip-based transmitter and receiver, hence it does not need additional hardware.  The idea is, Bob directly evaluates the QBER in $Z$- and $X$- basis from the public string after performing the time synchronization process. The estimated QBERs are then put into an optimization algorithm (based on gradient descent algorithm), which is running in Bob's laptop. The laptop then adjusts the applied voltages of PS in PC1 and PC2 using a programmable linear DC source. Taking advantage of gradient descent optimization algorithm, the voltages applied on PSs in PC1 and PC2 are tuned in real time to suppress QBER.  The process is continuously running until the estimated QBER is less a preset threshold.

\section{Experimental results}

\subsection{Characterization of components}

We first experimentally characterize each of components in the encoder chip. At a frequency of 50 MHz, the IM provides a static extinction ratio (ER) of approximately 27 dB and a dynamic ER of approximately 18 dB, which meets the requirement of one-decoy method~\cite{2018Rusca}.
The produced polarization states are analyzed with the decoder by Bob. The PCs are calibrated to  $Z$ ($X$) basis. We obtain an average polarization ER of $\sim$23 dB. The performance of the chip ensures realizing  a low-error, high-rate QKD system. 

The 3-dB bandwidth of CDM and PS, which are key parameters of evaluating the performance of the IM, Pol and PC, are measured. The 3-dB bandwidth of CDM is tested by measuring eye diagram and a value of approximately 10 GHz is obtained which is consistent with design. The 3-dB bandwidth of PSs is approximately 3 kHz. This value ensures the decoder a fast polarization tracking speed in field-buried and aerial fibers.

  The total loss of decoder chip is approximately 4.6 dB. The source of loss includes coupling-in and -out loss ($\sim$3 dB), on-chip device propagation loss ($\sim$1.6 dB). The MMI and waveguide bends contribute mostly to the propagation loss. The loss in the decoder chip reduces the probability of detecting photons, thus compromising the secure bit rate of QKD system.  

\subsection{Intrinsic QBER and Stability}
\begin{figure}[!ht]
	\centering
	\includegraphics[width=0.6\linewidth]{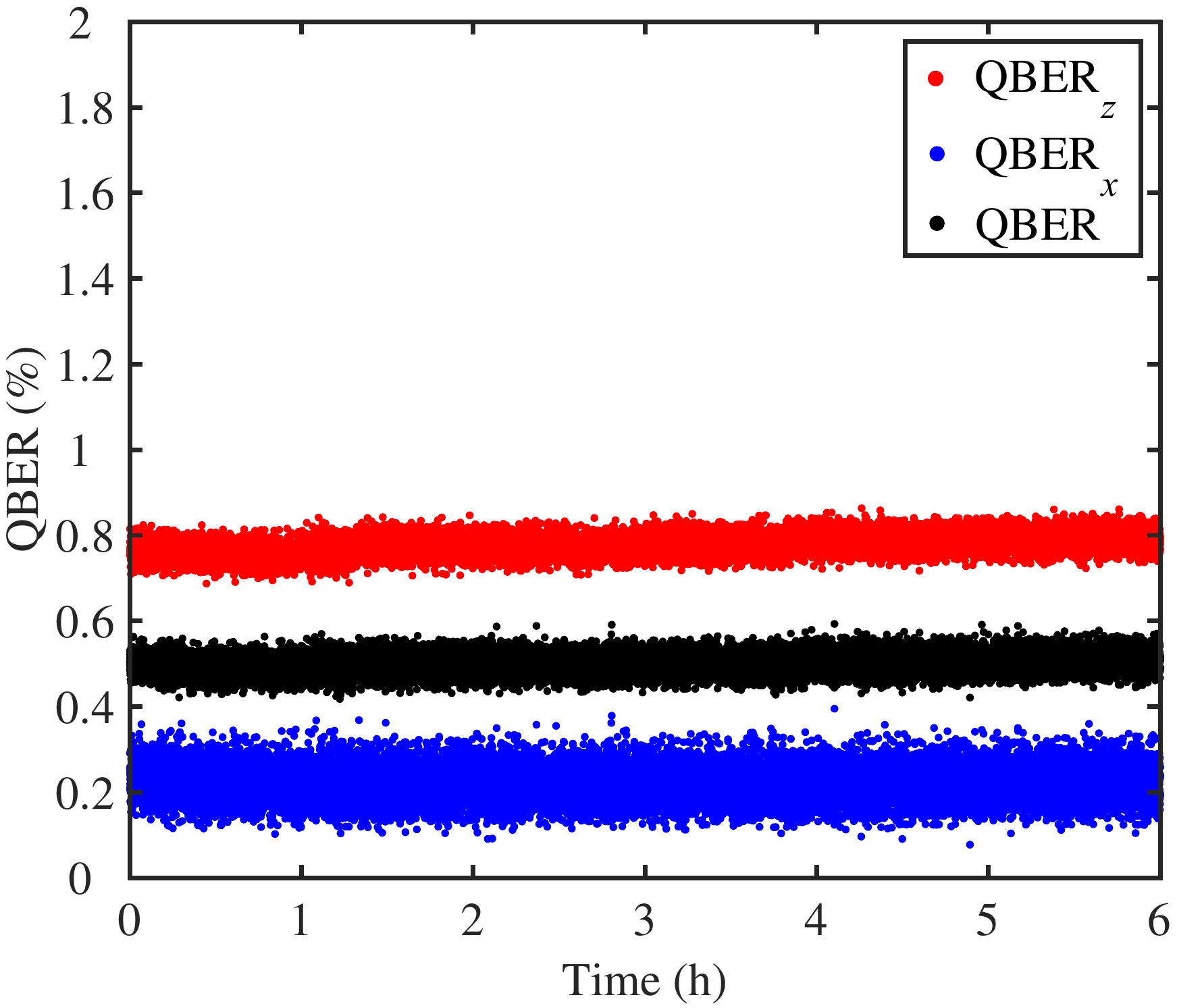}
	\caption{The quantum bit error on $Z~(X)$ basis of the system without active feedback over 6 h. The red (blue) points represent the quantum bit error in $Z~(X)$ basis. Each point  is refreshed every 1 second. }
	\label{QBERZX}
\end{figure}

We first report the intrinsic QBER and stability of our QKD setup, which gives a quantitatively first look  of the performance  of our setup. Furthermore, the characterization is crucial to predict the secure key rate under different channels or detector technologies. We first remove  the fiber spool and then send a  sequence of qubits  prepared using the method described in Sec.~\ref{qubit-method}.  The total length of periodic-correlation codes $L=5\times10^4$ and the ratio $M$ is set to 9:1.  The transmitted states are analyzed, including time synchronization performance as well as QBER estimation, using the measurement devices in Bob.   The intensity of signal $\mu$ is set to a constant value of 0.5. The experimental results are shown in Fig.~\ref{QBERZX}, which shows, in a 6-hour testing time, an average  QBER$_{z}$ is measured to be $0.78\pm 0.02\%$ for the $Z$ basis, while for $X$ basis, average QBER$_{x}$ is $0.22\pm 0.04\%$. the total of QBER for both two bases is $0.50\pm 0.02\%$. These results demonstrate the low intrinsic QBER and high stability of our setup.  Due to  different applied voltages on CDMs for the preparation of the four polarization states, the phase-dependent loss characteristics and saturation of the CDMs result in a discrepancy in the QBERs of $Z$ and $X$.  Since the data are contentiously collected, these results demonstrate the time synchronization method works well in a long-time run. 

\subsection{Polarization compensation of 50 km fiber spool}
To test the performance of polarization compensation method, we continuously run the system with a 50-km fiber spool channel. We exploit a fiber polarization scrambler, of which the applied voltage are step-by-step increased 1 V  every five minutes, to mimic the polarization rotation of photons in long fiber spool.  With our polarization feedback algorithm, we are able to compensate the polarization drifting to a low QBER within one minute.  For comparison, we also measure the QBERs without activating the polarization feedback.

\begin{figure}[!ht]
	\centering
	\includegraphics[width=0.6\linewidth]{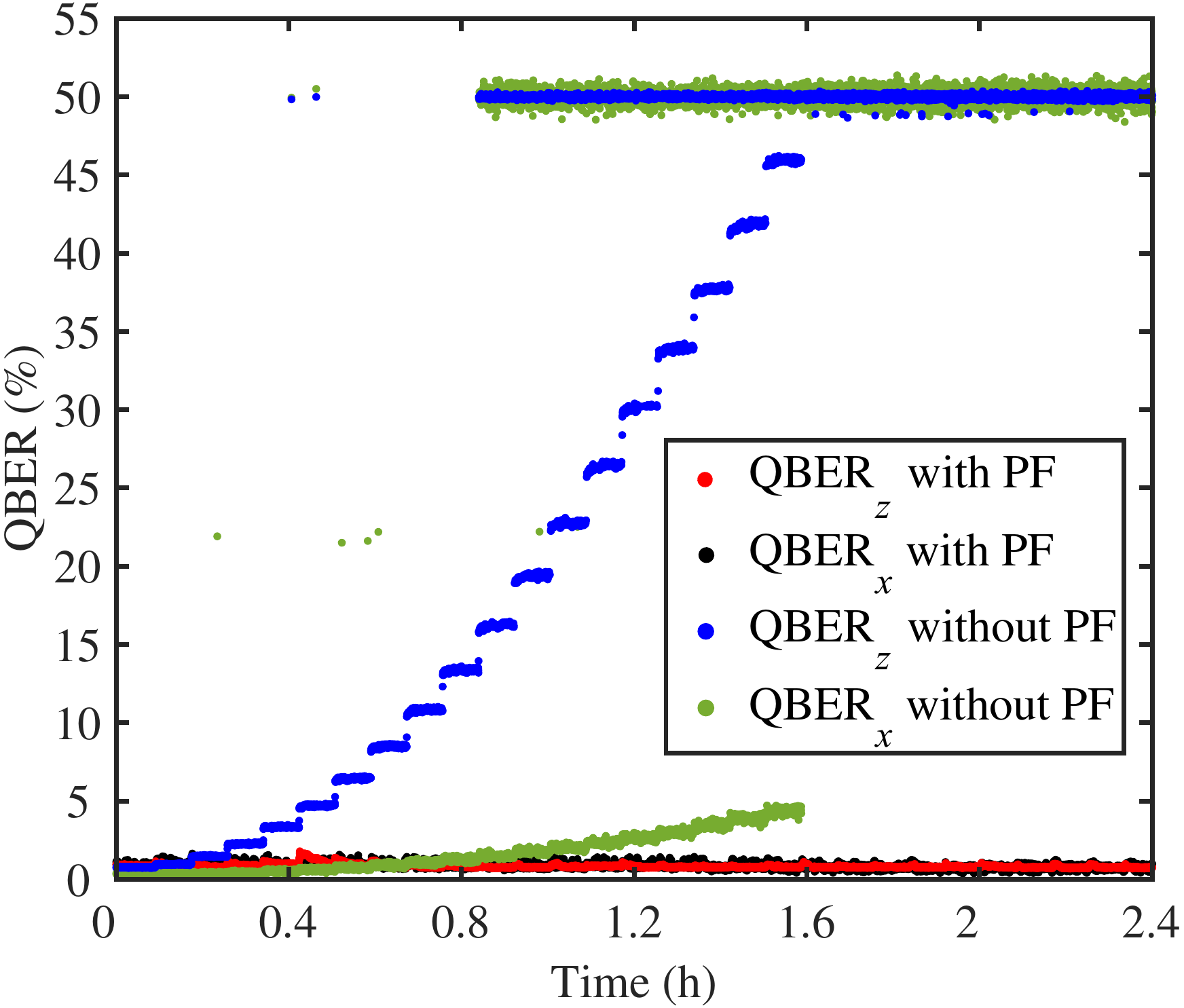}
	\caption{QBER measurements for a 50 km fiber channel over 2.4 h operation. During the measurement process,  a increased voltage at a step of 1 V is applied to the fiber scrambler every 5 minutes to mimic the polarization drift of the fiber channel. The orange~ (green) and blue ~(red) dots represent the measured QBER$_{x}$~(QBER$_{z}$) with and without polarization feedback (PF), respectively. With the PF, QBER$_{x}=0.86\pm 0.17\%$ and QBER$_{z}$ is $0.84\pm 0.13\%$.}
	\label{QBERZX-PF}
\end{figure}

The experimental results are shown in Fig.~\ref{QBERZX-PF}, It can be observed that, for 2.4-h continuous run,  due to the polarization drift introduced by the environmental distance and polarization scrambler, the QBERs for $Z$- and $X$-basis gradually increase to $50\%$ without active polarization feedback. In this stage, no secure key bits can be extracted. In contrast,  with our polarization feedback process, the QBERs keep stable,and stay at a low level of around 0.85\% for 2.4 h run.

\subsection{QKD secure key rate for different fiber channel lengths}

Using the described set-up, we perform a series of one-decoy QKD experiments with different fiber spool lengths of 50 km, 100 km, 150 km. 
To get a better performance, we perform a full optimization of the implementation parameters for the one decoy-state protocol~\cite{2022Li}. For example, the scenario of 100 km, the intensity of the signal states and decoy state are chosen to be $\mu=0.565$ and $\nu=0.143$, respectively. The probability of sending out the signal state $\mu$ and choosing the bases $Z$ are $P_\mu=0.798$ and $P_z=0.944$, respectively. At Bob's side, the probabilities of choosing the measurement basis in $Z$ and $X$ are permanently set to be equal due to the balanced MMI used in the decoder.

 For each distance, the total length of periodic-correlation codes is of $5\times10^4$ and the ratio $M$ is set to 9:1.  We continuously accumulate the detection numbers $n_z$ of approximately $10^7$ and perform the finite-key analysis using~\cite{2018Rusca} 

\begin{equation}\label{R_finite}
	\begin{split}
		R \leq [s_{{z}, 0}^{L} &+s_{{z}, 1}^{L}\left(1-h\left(\phi_{{z}}^{U}\right)\right)-\text{leak}_{\mathrm{EC}} \\
		&-6 \log _{2}\left(19 / \epsilon_{\mathrm{sec}}\right)-\log _{2}\left(2 / \epsilon_{\mathrm{cor}}\right)]/t ,           	
	\end{split}  
\end{equation}
where $t$ is the time duration of each acquisition, $s_{z,0}^{L} $ is the lower bound of the detector event received by Bob when Alice sends a vacuum state in  $Z$-basis, $s_{z,1}^{L} $ is  the lower bound of the detection event, given that Alice only sends a single-photon state in the $Z$-basis. $\phi _{z}^{U} $ is the upper bound of the phase error rate, $\text{leak}_{\mathrm{EC}}$ is the number of bits consumed for error correction, and $\epsilon_{sec}$ and $\epsilon_{cor}$ are the parameters used to evaluate secrecy and correctness, respectively. The $h\left ( x \right )$ denotes the Shannon binary information function.

\begin{figure}[h]
	\centering
	\includegraphics[width=0.6\linewidth]{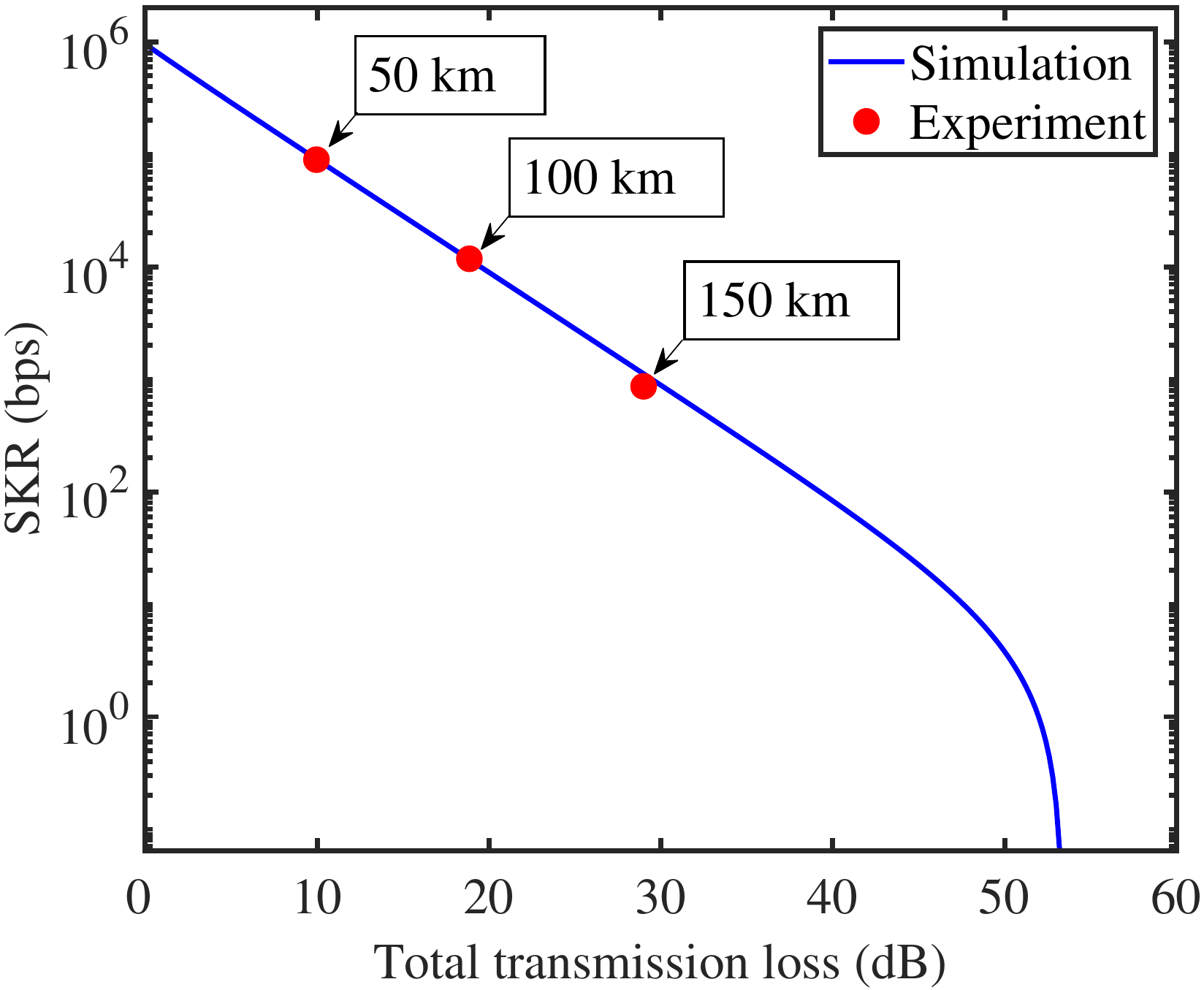}
	\caption{Secure key rates with different transmission loss. The blue line represent the simulation results based on our experimental parameters, and the red dots represent the experimental results. }
	\label{SKR}
\end{figure}

An overview of experimental parameters and performances for different distances are listed in Table~\ref{table_overview}. The experimental results are also plotted in Fig.~\ref{SKR}. It can been seen that, we successfully run QKD up to 150 km: with a continuous accumulated time of $3640.4~$s, a finite key rate of 866 bps is obtained. These results also demonstrate that our qubit-based synchronization and polarization compensation methods work well even in a high loss fiber channel and a long running time.
The detailed experimental results  are listed in~\ref{rawdate}.

\begin{table*}[!ht]
	\centering 
	\caption{ Overview of experimental parameters and performances for different distances. $L$ and Loss are the channel lengths and channel loss, respectively.  $\mu~(\nu)$ is the intensity of the signal (decoy) state, $n_{z}$ is the amount of data accumulated on the $Z$ basis, $t$ is the total accumulated time, $\tau _{b}$ is the period measured by Bob, QBER$_{z}$ is quantum bit error rates of $Z$ basis, $\phi _{z}^{U}$ is the upper bound of phase error rate, $s_{z,1}^{L}$ is the lower bound for measuring single-photon events in the $Z$ basis, and SKR is the final secure key rate.  }
	\resizebox{\linewidth}{!}{
		\begin{tabular}{ccccccccccc} 
			\hline\hline
			$L$ (km) & Loss (dB) &$\mu$ &$\nu$ & $n_{z}$ & $t$ (s) & $\tau _{b}$ (ns) & QBER$_{z}$ & $\phi _{z}^{U}$ & $s_{z,1}^{L} $ & SKR (bps)   
			\\ \hline
			50    & 9.957 & 0.568 & 0.144 & 10241262 & 42.3 & $20.0\pm5.4\times10^{-9}$ & $0.653\%$  & 0.0224  & 5280589 & $8.96\times10^4$   
			\\
			100    & 18.857 & 0.565 & 0.143 & 10021841 & 317.8 & $20.0\pm1.2\times10^{-8}$ & $0.764\%$  & 0.0177  & 5155932 & $1.18\times10^4$  
			\\
			150    & 28.992 & 0.564  & 0.142  &10078187 & 3640.4 & $20.0\pm1.9\times10^{-7}$ & $1.358\%$ & 0.0270  & 5317027 & $8.66\times10^2$
			
			\\ \hline\hline
		\end{tabular}
	}\centering \label{table_overview}
\end{table*}



\section{Discussion}

We have demonstrated a resource-efficient fully chip-based BB84 QKD scheme. The scheme implemented the time synchronization and polarization compensation using on-chip devices which generates also secure keys for quantum communication. The system can distill a finite-key secure bit rate of 866 bps over up to 150 km fiber spool.  Limited by driving circuit, which only supplies maximum peak-to-peak voltage $V_{pp}=10$ V at a repetition rate of 50 MHz, current system works at a repetition rate of 50 MHz and has a relatively low key rate. The obtained key rate can be further increased by using a state-of-the-art high speed modulator driving circuit~\cite{2020Liu-driving}.   The presented system can be further integrated with the laser based on wire-bounding or the hybrid integration technologies~\cite{2019Semenlaser,2019Agnesi}, which contributes to a compact chip-scale QKD transmitter.  Since CDMs are used to modulate the phase, it inevitably introduces the phase-dependent losses. To solve this problem, a polarization-loss-tolerant protocol~\cite{2018Lichenyang,2022Huang} can be incorporated into our work or novel phase modulation schemes provided in Ref.~\cite{2022Ye-chip,2017Sibson,2023Li-Chip} can be employed.     This work paves the way for low-cost, robust wafer-scale manufactured QKD system and provides a promising scheme for developing simple chip-based systems.

$Note~added$-During the preparation of this manuscript, the authors noticed that a related work which presented a high-speed integrated QKD system~\cite{2022Sax} has been released.  Although both works involve the chip-based QKD system, there are several differences. Firstly, in contrast to the use of a time-bin encoding scheme, our work employs a polarization encoding scheme, which does not require the maintenance of interferometric stability that would be necessary in time-bin encoding. Additionally, polarization encoding is also a favored choice for the deployment of quantum networks and free-space QKD, especially for ground-to-satellite QKD. 

Secondly, the receiver chip in Ref.~\cite{2022Sax} was fabricated on an aluminum borosilicate glass process platform, whereas we use CMOS-compatible silicon photonics technology, which has advantages of high-density integration and mature fabrication processing. 
Thirdly, Ref.~\cite{2022Sax} realized synchronization by using classical communication between two FPGAs, while our work uses a low-cost and efficient qubit synchronization approach.
 

\appendix
\begin{appendices}
 
\renewcommand{\thesection}{Appendix~\Alph{section}}

\section{Encoder and decoder design and fabrication}\label{chip-design}
The encoder chip has a device footprint of 3~mm $\times$ 6~mm. The device is fabricated on a high-resistance silicon-on-insulator  wafer with a 220 nm thick silicon layer and a 3 \textmu m thick buried oxide. The width of routing silicon waveguide was 450 nm.

The light was coupled in/out of the device via 1-dimensional and 2-dimensional grating coupler, respectively. The $2\times 2$ MMI couplers were designed as around 8 \textmu m $\times$ 60 \textmu m to get a nearly balanced splitting ratio. All the thermal-optical phase shifters (PSs) are identical with a length of 260 \textmu m, which efficiently result in a static extinction of around 28 dB when implemented in MZIs. Note that in the experiment, only one PS on one arm in each MZI was active, the other PS was designed for the compensation of 0.05 dB loss. 

All carrier depletion modulators (CDMs) are identical as well. The active CDMs have a length of about 3~mm. The doping concentration of the P and N type region is about $8\times 10^{17}$~cm$^{-3}$~ \cite{2018Li.Photon.Res.}. Intermediate P+ and N+ doping regions of concentrations of $3\times 10^{18}$~cm$^{-3}$ are added 220 nm away from the PN junction to reduce the series resistance and optical loss. The two PN junctions of CDMs are connected in series to form a push-pull configuration, which is beneficial to decreasing the capacitance and improving the impedance matching. Experimentally, the MZIs were biased with PSs, the CDMs inside MZIs were operated at high-speed with GSG traveling wave electrode. The estimated 3-dB bandwidth of the device was around 20 GHz. The half-wave voltage was smaller than 6 V. The Aluminum DC pad pitch is about 150 \textmu m. The pad area is about 80$\times$100 \textmu m$^2$.

The decoder chip has a device footprint of 1.6$\times$1.7 mm$^2$. The device is fabricated on a standard silicon-on-insulator wafer with a 220-nm silicon layer and a 3-\textmu m buried silica oxide layer. The width of the single-mode silicon waveguide was 450 nm.

The light is coupled into/out of the device via a spot-size converter (SSC) with a taper length of approximately 100 \textmu m and then split and converted into transverse electric (TE) modes using a compact polarization   splitter-rotator (PSR)~\cite{2021Chen}. The PSR comprises the following two functional parts: TM$_0$-TE$_\text{n}$ mode converter and TE$_\text{n}$-TE$_0$ mode converter. The extinction ratio of TM$_0$ and TE$_0$ mode of PSR are more than 17 dB over a wavelength range from 1520 to 1620 nm. The principle of the TM$_0$-TE$_\text{n}$ mode converter is the mode hybridization of the tapered rib waveguides, while the TE$_\text{n}$-TE$_0$ mode converter, acted as a polarization splitter rotator, is realized through the beam shaping method. The further details of operation principle can be found in Ref. \cite{2021Chen} and Ref. \cite{2016Chen}.   The variable optical attenuators (VOAs) function based on the forward carrier injection PIN junction, the value of attenuation is increasing with the applied voltage. The length of the VOA is approximately 200 \textmu m. The 1$\times$2 MMI and 2$\times$2 MMI couplers are designed to be approximately 4$\times$14 \textmu m$^2$ and 8$\times$60 \textmu m$^2$, respectively, to obtain a  balanced splitting ratio. All eight PSs are identical, with a length of 260 \textmu m, which efficiently results in a static extinction of approximately 28 dB when implemented in MZIs. Notably, in real-time, only one PS on one arm in each MZI was active and the other PS is designed for the compensation of a 0.05-dB loss. The pitch of the aluminum DC pad is approximately 150 \textmu m. The pad area is approximately 80$\times$100 \textmu m$^2$.

\begin{figure}[h]
	\centering
	\includegraphics[width=0.6\linewidth]{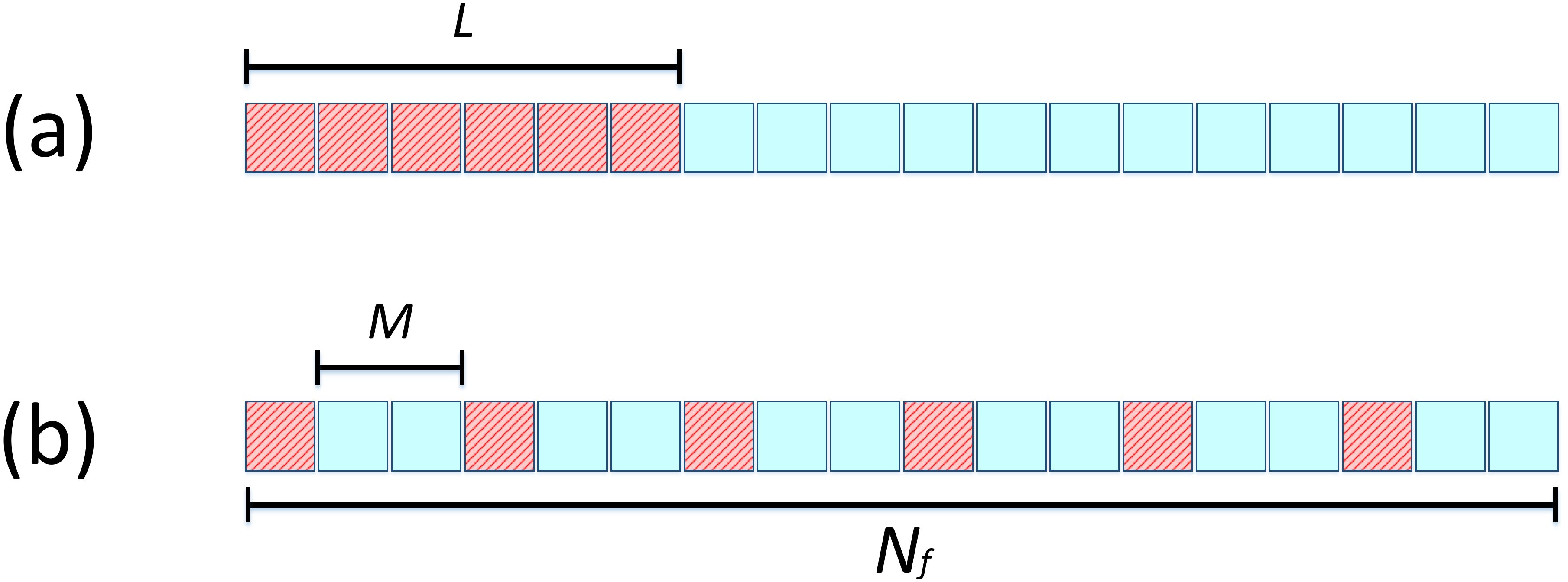}
	\caption{The way Alice sends a synchronization string. The red and blue squares represent single bits of the synchronization and random string, respectively. (a) Qubit4Sync. Alice first sends a synchronous string of length $L$ and then sends a random string for QKD. (b) Our method. Alice divided the synchronization string of length $L$ into $L$ single bits, and each bit is followed with $M$ random bits to form a block. $L$ blocks build the synchronization frame of length $N_f=(M+1)L$, which contains the complete synchronization string.}
	\label{fig_QDF}
\end{figure}

\section{Qubit-based synchronization method}\label{time-syn}

Here we describe our qubit-based synchronization method in detail. 
The main goal of the time synchronization is to recover the specific frequency of the qubits  arriving at the detectors and the time-offset of corresponding qubits between Alice and Bob. The main idea of our algorithm is similar to a recently proposed $Qubit4Sync$ algorithm~\cite{2020Calderaro} except for the way of placing the correlation codes, aiming at recovering the absolute time-offset. Suppose that Alice sends a qubit string which consists of several qubit frame with a length of $N_f$.  We define $L$ as the total length of the synchronization string  $s^A$  in each frame.  In the $Qubit4Sync$ algorithm, as shown in Fig.~\ref{fig_QDF} (a),  public periodic-correlation codes are encoded in the first $L$ states, in contrast, the codes with a length $L$ are divided into single bits and are then  periodically embed in prepared states in our method, as shown in Fig.~\ref{fig_QDF} (b). The detailed process is explained as follows.

As shown in Fig.~\ref{fig_QDF} (b),  Alice first generates  correlation codes $s^A$ of length $L$,  then divides them into $L$ single bits. Each bit is then connected with $M$ random bits.  These qubit strings are sent to Bob via a lossy channel. Bob performs time synchronization based on the received qubits. That is, Bob needs to determine the expected time of arrival $t^e_a$, which  can be expressed as 
\begin{equation}
	t^e_a=t_0+n_a\tau^B+\epsilon_{a}, \quad n_{a} \in {\mathbb{N}},
\end{equation}\label{Eq_te}
where $n_a$ is the order number of the sent qubit in Alice’s raw key, $a$ denotes the $a$-th detection, and $t_0$ is the initial time-offset, and $\epsilon_{a}$ is a  variable satisfies normal distribution, whose expectation value is zero, and variance is  $\sigma^2$.

We first describe how Bob  recovers the clock time $\tau^B$ from the received signal.  Bob performs a fast Fourier transform of the times-of-arrival signal. 
Similar to Ref.~\cite{2020Calderaro}, we sample the time of received signal with a sampling rate of $4/\tau^{A}$, where  $\tau^A$ as is the clock time of Alice for sending the light pulse. The number of samples for the fast Fourier transform
is set to $N_{s}=10^6$. At this stage, Bob gets a primary estimation $\tau^B_0$ of $\tau^B$ with an error of approximately  $4\tau^{A}/N_s$. To get a more accurate value, the sampled signals are then fed into the least trimmed squares algorithm to get an accurate clock time which satisfies 

\begin{equation}
	\frac{1}{D} \sum_{b=1}^{D}\left|\mathcal{E}_{a}^{I}(b)\right|^{2} \simeq \sigma^{2}. 
\end{equation}\label{Eq_period}
Here $\mathcal{E}_{a}=t_{a}^{m}-t_{a}^{e}$  indicates the time-interval error,  $t^m_a$ is the measured time of arrival with $a \geq 1$, in Bob site,  $\mathcal{E}^I_{a}(b)=\mathcal{E}_{a+b}-\mathcal{E}_{a}$ denotes  the time-interval error between two different detections $a$ and $a+b$. $\mathcal{E}^I_{a}(b)$ can help Bob measure the accuracy of the period recovery. 
\begin{figure}[t!]
	\centering
	\includegraphics[width=0.9\linewidth]{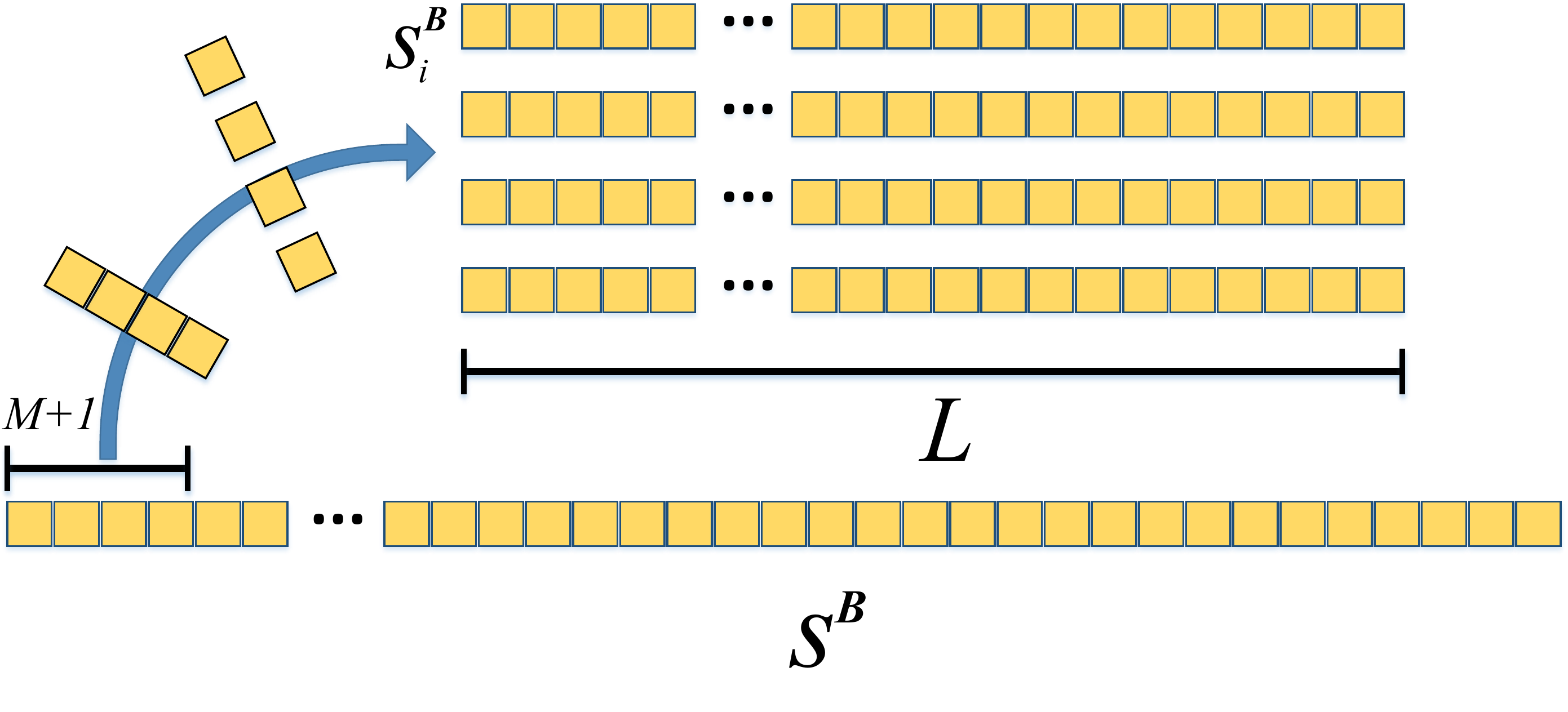}
	\caption{Schematic of reverse-processing $s^B$. Bob's string  $s^B$ are divided into $L$ blocks of length $(M+1)$ and reshaped into $(M+1)\times L$ matrix. Each row of the matrix reconstructs a bit string   $s^B_i$.}
	\label{fig_SYSnetwork}
\end{figure}

Now we describe how to calculate the time-offset $t_0$ from the detections. It should be known that once $\tau^B$ have be identified, Bob has  the correct index difference $n_{a+1}-n_a$. Due to the losses of  channel, it is unlikely that the first photon in pulse is detected by Bob. So Bob makes a first guess $t^A_0$ about the value of  $t_0$, where $t^A_0$ is regarded as Bob's first detection as well as the first pulse sent by Alice.

In order to precisely determine $t_0$, Bob should calculate the correlation between the received signal and the public synchronization string $s^A$. 
The synchronization string is only encoded on the $Z$ basis and we assign the values $+1$ and $-1$ to the $H$ and $V$ state, respectively. Once the period $\tau^B$ is determined, Bob makes a first guess of $t_0$, then extracts  the detection can generate a string $s^B$  with values  $+1$, $-1$ and $0$, in which the value of $0$ denotes no detection or a detection in $X$ basis. Note that $s^B$ has a length of $N_f$. Evidently, $s^B$ must contain the detections of a complete synchronization string of length $L$. 
\begin{table*}[b!]
	\centering 
	\caption{ Experimental parameters and results. 
		$L$ and Loss are the channel lengths and channel losses, respectively.  $\mu~(\nu)$ is the intensity of the signal (decoy) state, $P_{\mu}$ ($P_{\nu}$) is the selection probability of the signal (decoy) state, $P_{z}$ ($P_{x}$) is the probability of choosing $Z~(X)$ basis, $n_{z}$ is the amount of data accumulated on the $Z$ basis, $t$ is the total accumulated time of measurement, Ratio denotes the ratio between the number of qubits for synchronization and that of random polarization states of four BB84 polarizations. $\tau _{b}$ is the period measured by Bob, $n_{z,\mu}$ ($n_{x,\mu}$) is the raw count measured under  $Z~(X)$ basis for the signal state, $n_{z,\nu}$ ($n_{x,\nu}$) is the raw count measured for the decoy state under  $Z~(X)$ basis, $m_{z,\mu}$ ($m_{x,\mu}$) is the error count measured under  $Z~(X)$ basis for the signal state, $m_{z,\nu}$ ($m_{x,\nu}$) is the error count measured for the decoy state under  $Z~(X)$ basis, QBER$_{z}$ is quantum bit error rates of $Z$ basis, $\phi _{z}^{U}$ is the upper bound of phase error rate, $s_{z,1}^{L}$ is the lower bound for measuring single-photon events in the $Z$-basis, $l$ is the secret key length, and SKR is the final secure key rate.}
	\resizebox{\linewidth}{!}{
		\begin{tabular}{ccccccccccccc} 
			
			\hline\hline
			$L$ (km) & Loss (dB) &$\mu$ &$\nu$ &$P_{\mu}$ &$P_{\nu}$ & $P_{z}$ &$P_{x}$ &Ratio& $n_{z}$ & $t$ (s) & $\tau _{b}$ (ns) &QBER$_z$  
			\\ \hline
			50    & 9.957 & 0.568 & 0.144 &0.799 & 0.201 & 0.944 & 0.056 & 1:9 & 10241262 & 42.3 & $20.0\pm5.4\times10^{-9}$ &$0.653\%$  
			\\
			100    & 18.857 & 0.565 & 0.143 &0.798 & 0.202 & 0.944 & 0.056 & 1:9 & 10021841 & 317.8 & $20.0\pm1.2\times10^{-8}$ & $0.764\%$
			\\
			150    & 28.992 & 0.564  & 0.142 & 0.798  & 0.202 & 0.944 &0.056 & 1:9 &10078187 & 3640.4 & $20.0\pm1.9\times10^{-7}$ &$1.358\%$
			
			\\ \hline\hline
			\\ \hline\hline
			
			$L$ (km) & $n_{z,\mu}$ & $m_{z,\mu}$ & $n_{x,\mu}$ & $m_{x,\mu}$ & $n_{z,\nu}$ & $m_{z,\nu}$ & $n_{x,\nu}$ & $m_{x,\nu}$ &  $\phi _{z}^{u}$ & $s_{z,1}^{l} $ &$l$ & $SKR$(bps) 
			\\ \hline
			50     &  9628161 & 62566 & 575986 & 3379 & 613101 & 4387 & 35863 & 236&  0.0224  & 5280589 &3787713  & $8.96\times10^4$ 
			\\
			100    &  9419400 & 70873 & 557703 & 2576 & 602441 & 5644 & 35140 & 200&  0.0177  & 5155932 &3742736  & $1.18\times10^4$  
			\\
			150    &  9456469  & 126891 & 577883 & 4653 & 621718 & 9973 & 37593 & 334&  0.0270  & 5317027 &3152614 & $8.66\times10^2$
			
			\\ \hline\hline
		\end{tabular}
	}
	\label{table_rawdate}
\end{table*}

As shown in Fig.~\ref{fig_SYSnetwork}, Bob can reversely processes $s^B$ according to the frame structure shown in Fig.~\ref{fig_QDF} (b). Bob first separates the detection string $s^B$ into $L$ blocks of length $M+1$ and then fits them  into an $(M+1)\times L$ matrix. Each row of the matrix constructs a string  $s^B_i$ with a length of $L$, where $i \in \{1,2,...,M+1\}$. By calculating the cross-correlation between the $s^A$ and $s^B_i$, Bob can determine the optimal detection string $s^B_{i_{opt}}$ which offers a maximum value of cross-correlation.  The cross-correlation calculation between $s^B_{i_{opt}}$ and $s^A$ can be accelerated using the matrix manipulation  described in Ref.~\cite{2020Calderaro}. With the optimal value of the cross-correlation, an accurate time-offset between Alice and Bob can be defined as illustrated in Ref.~\cite{2020Calderaro}.
 For a good offset recovery, the length of $s^A$ should satisfy the following inequality: $\sqrt{L\eta}\geq10$, where $\eta$ is the total transmittance.  When the system has a very high channel loss, we also can overcome the loss by repeating synchronization string detections. This avoids increasing the length $L$ of the synchronization string.  The efficiency of QKD can further improved by using a shorter length of $s^A$; however,  to ensure the correct execution of the synchronization algorithm, it cannot be indefinitely diluted. This is determined by two factors: the time duration of system frequency stability and the total transmission loss.

\section{Detailed experimental results}\label{rawdate}
Table~\ref{table_rawdate} shows the detailed experimental results.

\end{appendices}

\begin{backmatter}
	\bmsection{Funding} This study was supported by the National Natural Science Foundation of China (Nos. 62171144 and 62031024), the Guangxi Science Foundation (No.2021GXNSFAA220011), and the Open Fund of IPOC (BUPT) (No. IPOC2021A02).
\end{backmatter}

\begin{backmatter}
	\bmsection{Acknowledgments} We thank Shizhuo Li for drawing the diagram of the chip.
\end{backmatter}

\begin{backmatter}
	\bmsection{Author contributions statement} K.W. design and developed the experiment, X.Hu, X.Hua and X.X developed the device design and fabrication. K.W., Y.D., Z.Z, C.H. and Y.C. implemented the experiment, and evaluated the data. K.W. and X.X. supervised the
	research and experiment. All authors contributed to the writing of the manuscript.
	K.W. and X.Hu contributed equally to this paper.
\end{backmatter}

\begin{backmatter}
	\bmsection{Disclosures} The authors declare no conflicts of interest.
\end{backmatter}

\begin{backmatter}
	\bmsection{Data Availability} Data underlying the results presented in this paper are not publicly available at this
	time but may be obtained from the authors upon reasonable request.
\end{backmatter}


\bibliography{Chip-BB84-Syc}

\end{document}